# Drift-compensated Low-noise Frequency Synthesis Based on a cryoCSO for the KRISS-F1

Myoung-Sun Heo, Sang Eon Park, Won-Kyu Lee, Sang-Bum Lee, Hyun-Gue Hong, Taeg Yong Kwon, Chang Yong Park, Dai-Hyuk Yu, G. Santarelli, Ashby Hilton, Andre N. Luiten and John G. Hartnett

***Abstract*—In this paper we report on the implementation and stability analysis of a drift-compensated frequency synthesizer from a cryogenic sapphire oscillator (CSO) designed for a Cs/Rb atomic fountain clock. The synthesizer has two microwave outputs of 7 GHz and 9 GHz for Rb and Cs atom interrogation, respectively. The short-term stability of these microwave signals, measured using an optical frequency comb locked to an ultra-stable laser, is better than $5\times10^{-15}$ at an averaging time of 1 s. We demonstrate that the short-term stability of the synthesizer is lower than the quantum projection noise limit of the Cs fountain clock, KRISS-F1(Cs) by measuring the short-term stability of the fountain with varying trapped atom number. The stability of the fountain at 1-s averaging time reaches $2.5\times10^{-14}$ at the highest atom number in the experiment when the synthesizer is used as an interrogation oscillator of the fountain. In order to compensate the frequency drift of the CSO, the output frequency of a waveform generator in the synthesis chain is ramped linearly. By doing this, the stability of the synthesizer at an average time of one hour reaches a level of $10^{-16}$ which is measured with the fountain clock.**

***Index Terms*—Cryogenic sapphire oscillator (CSO), optical frequency comb, frequency stability, frequency synthesizer, hydrogen maser, stable lasers, quantum projection noise.**

## I. Introduction

THE most accurate realization of the International System (SI) second has now been made by the atomic fountain frequency standards, some of which reached an uncertainty at a level of $10^{-16}$ [1]–[5]. To this end, an interrogation oscillator, a microwave signal for interrogation of the atomic resonance, needs to meet stability requirements at both short and long measurement time. Good short-term stability is essential to reduce the effect of the phase noise of the local oscillator on the pulsed operation of the clock [6], [7] and allows fountain clocks to operate in the quantum projection noise limited regime (QPN) [8], which is at a level of about $10^{-14}$ for $10^6$ atoms. In addition, better short-term stability greatly facilitates the accuracy evaluation by dramatically shortening the time needed to reach the target stability. Long-term frequency stability at a level of a few parts in $10^{-16}$ or better is necessary to reach the target uncertainty of $10^{-16}$.

As an interrogation oscillator for the Cs atomic fountain clock in the Korea Research Institute of Standards and Science, KRISS-F1(Cs), we used a hydrogen maser (H-maser) filtered by a low noise BVA quartz oscillator. This approach meets the long-term stability requirements but limits the short-term stability of the fountain clock to about $1\times10^{-13}$ at a 1-s averaging time, dominated by the phase noise of the interrogation oscillator. Recently we introduced a cryogenic sapphire oscillator using an ultralow vibration pulse-tube cryocooler (cryoCSO) developed by Hartnett's group [9]. The stability of this cryoCSO was measured using another cryoCSO before shipment to the KRISS lab, as described in [10] (curve 4 in Fig. 6). The short-term stability of the cryoCSO was measured to be as low as $2\times10^{-15}$ at 1-s averaging time with a long-term frequency drift of about $5\times10^{-14}$/day. The output frequency is at 11.200044 GHz, so we designed a stable frequency synthesizer for the fountain clock as briefly described in [11]. The short-term stability of its microwave output signals, which was measured by comparison with an ultra-stable laser via an optical frequency comb, is better than $5\times10^{-15}$ over an averaging time of 1 s. We confirmed that the short-term stability of the KRISS-F1(Cs) operating with this synthesizer is QPN limited for a wide range of atoms. The long-term frequency drift is usually compensated by a loose phase-lock to a H-maser with a large time constant (~ 1000 s) [3], [12]–[18]. In this paper, since this drift is almost linear, we cancelled the linear frequency drift by applying a linear frequency ramp to the synthesizer. With this passive compensation, the long-term frequency stability reached $4\times10^{-16}$ at a day.

## II. Design of the Frequency Synthesizer

We built a frequency synthesizer with the cryoCSO as depicted in Fig. 1. The output frequency of the cryoCSO is about 11.200044 GHz. The frequency drift of the cryoCSO was measured to be almost linear with the rate of 6.2 nHz/s. For the synthesis of exact integer frequencies, the small offset frequency 44 kHz was removed using an IQ mixer [19] and two digital waveform generators (WGs) as shown in the shaded grey region in Fig. 1. Because the linear drift rate is

Manuscript received. M.-S. Heo, S.E. Park, W.-K. Lee, S.-B. Lee, H.-G. Hong, T.Y, Kwon, C.Y. Park and D.-H. Yu are supported by the Korea Research Institute of Standards and Science under the project 'Establishment of National Physical Measurement Standards and Improvements of Calibration/Measurement Capability', grant 16011007. A. Hilton, A.N. Luiten and J. G. Hartnett are supported by the Australian Research Council for funding through grants LP110200142 and LE130100163.

M.-S. Heo, S.E. Park, W.-K. Lee, S.-B. Lee, H.-G. Hong, T.Y, Kwon, C.Y. Park and D.-H. Yu are with the Center for Time and frequency, Division of Physical Metrology, Korea Research Institute of Standards and Science, Daejeon 341113 Korea (e-mail: parkse@kriss.re.kr).

S.E. Park, W.-K. Lee and C.Y. Park are with the Science of Measurements, University of Science and Technology (UST), Daejeon 34114 Korea.

G. Santarelli is with the Laboratoire Photonique Numérique et Nanoscience (LP2N), UMR 5298, CNRS-IOGS-Université de Bordeaux, 33400 Talence, France.

A. Hilton, A.N. Luiten and J. G. Hartnett are with the Institute of Photonics and Advanced Sensing (IPAS), School of Chemistry and Physics, University of Adelaide, South Australia, Australia.

too small to be compensated with a single WG (1μHz frequency resolution), two WGs were used. We could compensate the linear drift by ramping 10-MHz output frequency from another WG (WG2) which is used as the reference frequency for the WG1. The ramping rate should be $1.4\times10^{-13}$/s = 6.2 nHz/s/44 kHz, so we changed the output frequency 10 MHz of the WG2 by 1 μHz every 710 ms. A fine adjustment of this 710 ms interval was made after a few days of observation of the frequency drift which is described in section IV.

After the IQ mixer, microwave dividers and mixers were combined to obtain microwave signals at 9 GHz for Cs clocks and 7 GHz for Rb clocks. To minimize the excess phase noise we selected ultra-low-phase-noise microwave dividers and amplifiers made by Analog Devices, as described in Fig. 1. To reject any unwanted sidebands, narrow bandpass filters (200 MHz/150 MHz 3-dB bandwidth for 9 GHz/7 GHz, K&L microwave) were inserted just before isolators at the output port of 9 GHz and 7 GHz. The filter packages are thermoelectrically stabilized to minimize temperature induced phase variations. The auxiliary synthesized radiofrequency outputs (5/10/100/800 MHz) are used for reference to various synthesizers or for comparison with other atomic clocks.

## III. Short-term Stability

The short-term stability of our cryoCSO was previously measured using another cryoCSO by Hartnett's group as seen in the dashed curve in Fig. 3. That approach is not feasible in our lab. We instead used the ultra-stable 578-nm clock laser developed for an Yb optical lattice clock and an optical frequency comb for stability analysis [13], [20]. In the later part of this section, using the 9 GHz signal from the synthesizer we operated the KRISS-F1(Cs) and verified that the short-term stability is not dominated by the noise of the interrogation oscillator, but instead by QPN.

### *A. Measurement with the stable optical clock laser*

We have developed an ultra-stable 578-nm clock laser for an Yb optical lattice clock [21]. It is locked to a high finesse optical cavity, made of ultralow thermal expansion glass. Its frequency stability (Allan deviation) can be measured either by comparison with another independent clock laser or by direct measurement using Yb atoms in the optical lattice. Its short-term stability is about $2\times10^{-15}$ for a 1-s averaging time.

Fig. 2 describes the measurement setup. To compare the optical laser frequency with microwave, we used an erbium-doped fiber frequency comb (FC1500, Menlo systems) centered at 1550 nm with a repetition rate of 250 MHz. We transferred the stability of the clock laser into the microwave frequency region using high-bandwidth (~300 kHz) phase-locking of the frequency comb to the clock laser with a built-in electro-optic modulator (EOM), along with the carrier-envelope offset frequency ($f_{ceo}$) control using feedback to the pump laser current. The out-coupled comb is amplified and fed into a low-noise photodetector (DSC50S, Discovery). Its output signal is mixed with the output signals of the synthesizer or the cryoCSO. By doing this, the down-converted outputs of the synthesizer can be used for stability measurements using the phase detector (5125A phase-noise test set, Microsemi) which has an input frequency range from 1 MHz to 400 MHz. The repetition rate $f_{rep}$ is adjusted to 249.775 MHz, so the down-converted frequencies of 7 GHz, 9 GHz and 11.2 GHz are 6.3 MHz, 8.1 MHz and 210 MHz, respectively.

Before making stability measurements using the setup in Fig.2, we first needed to verify that the 100 MHz reference to the 5125A test set has sufficiently low noise to measure the stability of the microwave signals at the level of $10^{-15}$ or below. Because the microwave signals are down-converted by 3 orders of magnitude, the stability of the 100 MHz reference needs only to be below $10^{-12}$. This was verified by comparison with 250 MHz output from the comb. The inset of Fig. 3 shows that the instability of the 100 MHz signal is below $10^{-14}$ up to 100-s averaging time.

Fig. 3 shows the stability for the 7 GHz and 9 GHz signals from the synthesizer and that of the 11.2 GHz signal from the cryoCSO. It is noted that these are combined stability of microwave signals generated both from the frequency comb and the synthesizer. The stability of the 11.2 GHz output directly from the cryoCSO is $4\times10^{-15}$ at 1 s, which is worse than that determined by Hartnett's group using another cryoCSO (see the dashed line in Fig. 3). This means that the short-term stability measurement is limited by the transferred stability of the frequency comb from the clock laser. Besides the electronic noise from the photodetection, we infer that our fiber frequency comb itself contributes to the instability from the fact that the measured stability at 1 s of the $f_{ceo}$ is at the level of 1 Hz, which corresponds to $4\times10^{-15}$ against 259 THz (half the clock laser frequency).

From the results of the stability measurements, we conclude that the synthesized microwave signals have a short-term stability of ~ $5\times10^{-15}$ at 1 s, which is sufficient to be used for an atomic fountain clock, and the stability of the output signals from the synthesizer is not degraded by more than a factor of 3 in comparison with that of the cryoCSO, assuming it still has the same frequency stability measured by Hartnett's group as shown by the dashed line in Fig. 3.

### *B. QPN-limited operation of the atomic fountain clock*

Since the short-term stability of the synthesized microwave signals are below $10^{-14}$, we can expect that an atomic fountain clock will operate in the QPN limit when these microwave signals are used as an interrogation oscillator. [22]

Fig. 4 shows the frequency synthesis chain of an interrogation oscillator for the KRISS-F1(Cs). The 9 GHz and 100 MHz output of the cryoCSO synthesizer are mixed with each other to make the 9.1 GHz microwave. The 9.1 GHz microwave signal is mixed again with a 92.63177 MHz output of a direct digital synthesizer (DDS), which is referenced to the 800 MHz signal from the cryoCSO synthesizer. To avoid aliasing in the DDS, the 800 MHz signal is used as the reference frequency of the DDS so that the Nyquist frequency of the DDS is much higher than the output frequency of about 92 MHz. Finally, the synthesized microwave of 9.19263177 GHz is sent to the atomic fountain. For pulsed microwave operation in the fountain, an interferometer switch [23] is used. The microwave signal delivered to the fountain is switched by switching the 100 MHz signal as shown in Fig. 4. The frequency of the microwave sent to the fountain for Ramsey interrogation is tuned by controlling the DDS output frequency.

To confirm that the stability of the microwave signal is better than the QPN limit in the fountain, the short-term stability of the fountain was measured with varying the trapped atom number as shown in Fig. 5. In the QPN limit, the stability at 1-s averaging time improves with increasing number of detected atoms [8] satisfying the following relation (1).

$$\sigma(\tau = 1\,\mathrm{s}) \propto N^{-1/2} \quad (1),$$

where $N$ is the number of detected atoms. Fig. 5 shows that the dependence of the stability on the number of detected atoms is in good agreement with (1). This result shows that the short-term stability of the fountain is not affected by the stability of the output signals from the synthesizer. Therefore, we confirm that the short-term stability of the synthesized microwave signal is better than the QPN limit of our atomic fountain. Fig. 5 shows the frequency stability improves to $2.5\times10^{-14}$ for the highest number of detected atoms.

## IV. Long-term Stability and Drift compensation

The long-term stability of the synthesizer was measured by comparing the synthesized 9 GHz with the KRISS-F1(Cs) using the setup described in Fig. 4. The synthesizer has a linear frequency drift of $4.8\times10^{-14}$/day coming from the cryoCSO (curve 1 in Fig. 6). This drift can make it difficult to carry out some of essential measurements for accuracy evaluation of the fountain clock if they take longer than a few thousands of seconds to reach targeted uncertainty, because the drift takes effect after an hour as seen in the curve 1 in Fig. 6. One example of these measurements is the evaluation of the sensitivity of the distributed cavity phase shift, which demands interleaved measurements for various mechanical tilts of the fountain clock [24]. In case of successive adjustments of two different tilts, total duration of measurements for a single tilt should be less than half an hour. This is challenging because we need to mechanically change the tilt every half an hour or less. Because the frequency drift has been quite linear, we decided to compensate passively the linear part of the drift within the frequency synthesizer setup as shown in Fig. 1, instead of active phase-locking to a H-maser.

Fig. 6 shows the frequency stability of the synthesized 9 GHz signal measured by comparing it with the KRISS-F1(Cs), and that of the cryoCSO itself measured with another cryoCSO by Hartnett's group before being shipped to our lab. As seen in the curves 1-3 in Fig. 6 the stability up to 1000 s is governed by the stability of the KRISS-F1(Cs). After 1000s, the frequency drift from the cryoCSO becomes dominant. This drift measurement agrees well with the measurement made using another cryoCSO (curve 4 in Fig. 6). We first tried compensation by ramping the frequency of the WG2 in Fig. 1 by 1 μHz frequency steps every 710 ms estimated from the curve 1. From the curve 2, it was found that there still existed the linear frequency drift due to an error in the estimation of the time interval from the curve 1. So we adjusted the interval by about 50 ms. After the second compensation, the long-term-stability could reach at a level of $7\times10^{-16}$ within an hour.

## V. Conclusion

We designed and implemented a frequency synthesizer using a cryoCSO for use as an ultra-stable interrogation oscillator of a Cs/Rb atomic fountain clock. The short-term stability of the microwave signals from the synthesizer was about $5\times10^{-15}$ at 1-s averaging time. With this ultralow noise microwave local oscillator, we have demonstrated the QPN-limited operation of our Cs fountain clock, KRISS-F1(Cs). From this result, we are assured that the short-term stability of the microwave synthesizer is below the QPN limit of our fountain. For the accuracy evaluation of the fountain with $10^{-16}$ uncertainty level, the long-term frequency drift of the cryoCSO was compensated by adjusting an offset frequency from a WG deployed in the synthesizer. The resultant long-term stability of the synthesized microwave signals, as measured by the fountain clock, can reach a level of a few parts in$10^{-16}$.

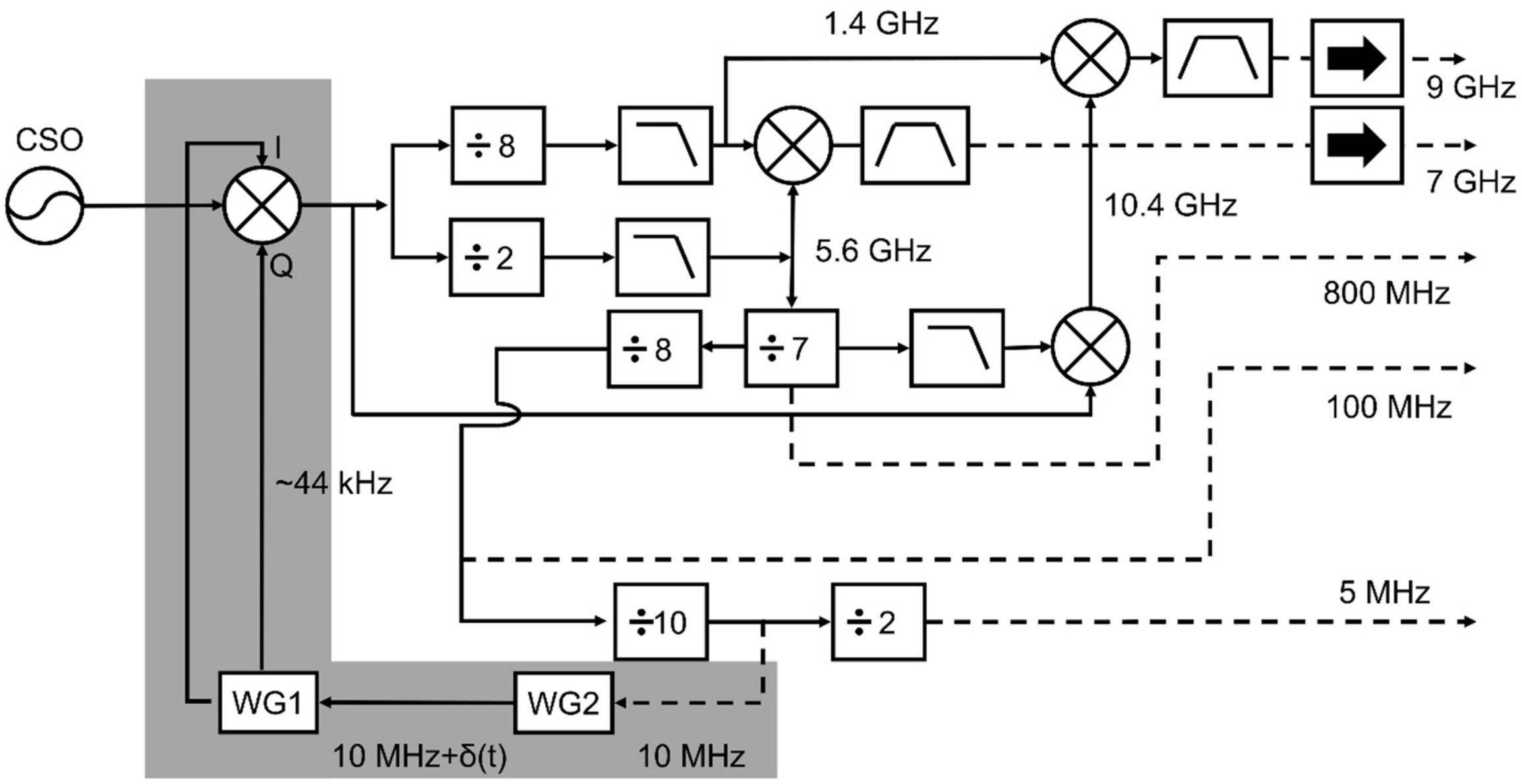

Fig. 1. Schematic diagram of the frequency synthesizer. Not all the attenuators, filters, amplifiers and isolators are shown. The first mixer in the grey region is an in-phase/quadrature (IQ) mixer (IQ-0714, Marki microwave). It was used for both the small offset frequency cancellation and frequency drift compensation. WGs are waveform generators (AFG3022B, Tektronix). The important low-phase-noise components used are the Analog Devices dividers HMC-C005 (÷2), HMC-C007 (÷8), and the programmable divider HMC705LP4 (÷7), the Analog Devices HMC-C050 amplifier, the Holzworth HX4210 divider (÷10), and the Wenzel LNFD-2-10-13-1-13 divider (÷2). The WG1 reference frequency 10 MHz from the WG2 is remotely controlled, so that it is ramped linearly. Details are described in the text.

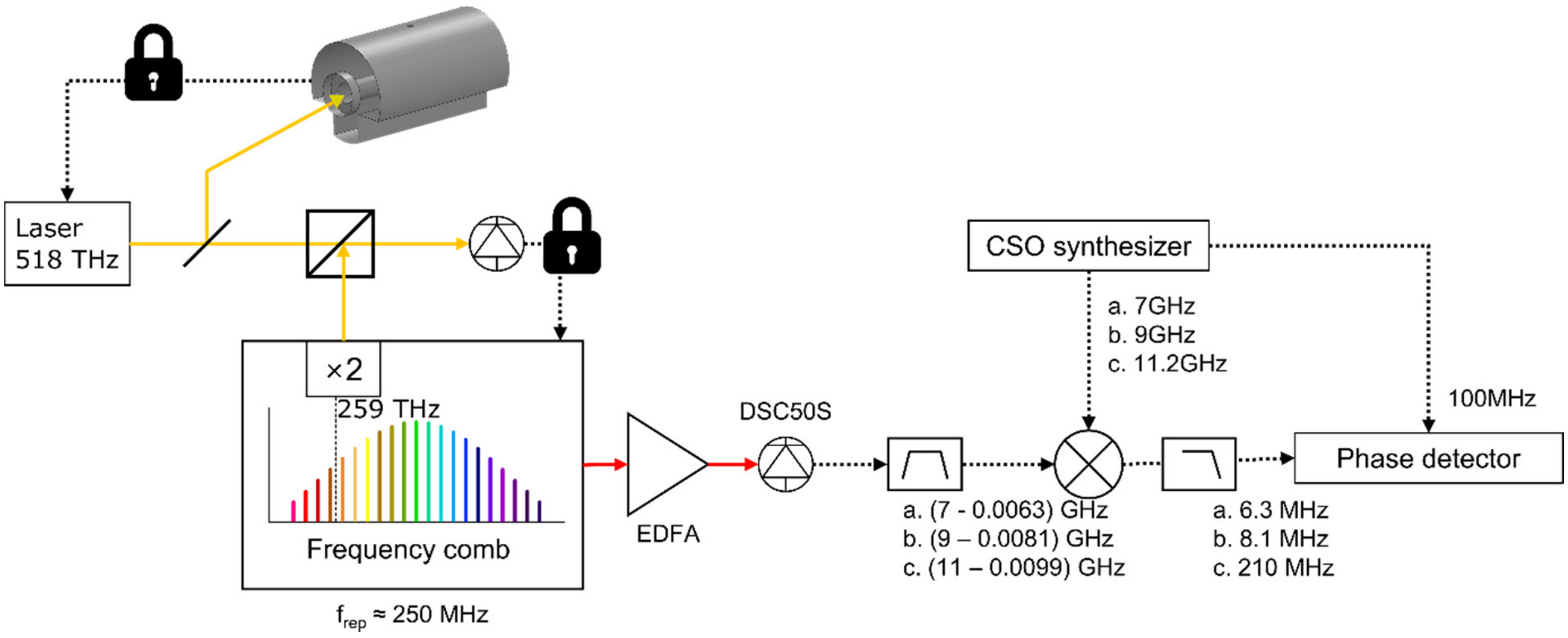

Fig. 2. Setup for the short-term stability measurements. Output signals from the synthesizer and the cryoCSO are down-converted by mixing with filtered beat signals from the frequency comb which is locked to the optical reference. The frequency stability of down-converted signals was measured using a phase detector, a Microsemi 5125A phase noise test set, which outputs phase difference data in units of cycles of the input signal. We used the 100 MHz signals from the synthesizer as a reference.

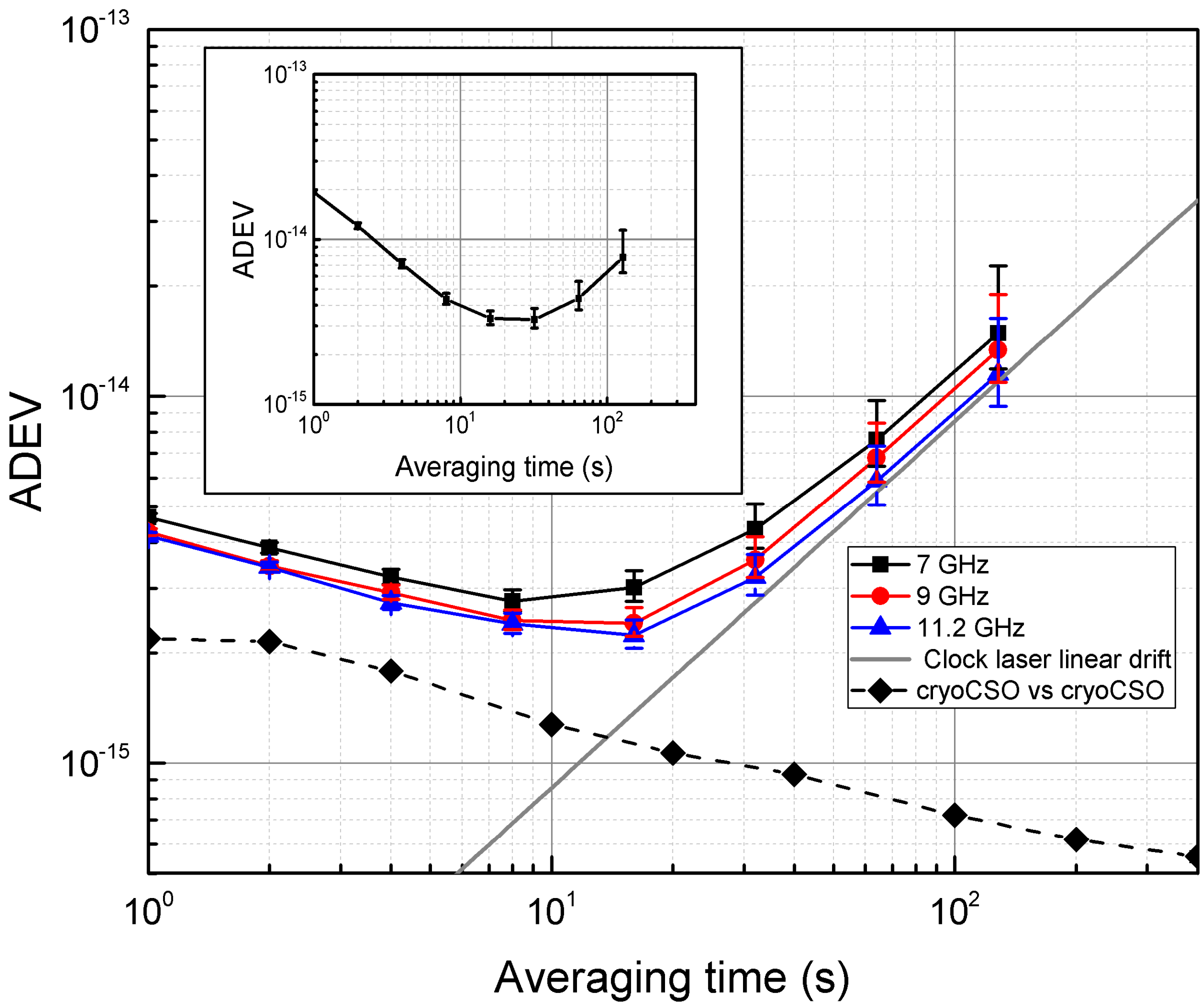

Fig. 3. Fractional frequency stability of the output signals from the synthesizer and the CSO. The noise equivalent bandwidth of the phase detector (5125A) is set to 0.5 Hz. The Allan deviation (ADEV) is computed from the phase difference data. The frequency drift after 10 s comes from that of the clock laser whose drift is depicted in the grey solid line. The dashed line shows the stability measurement using another cryoCSO. Small error bars are not visible. The inset is the stability of 100 MHz from the synthesizer.

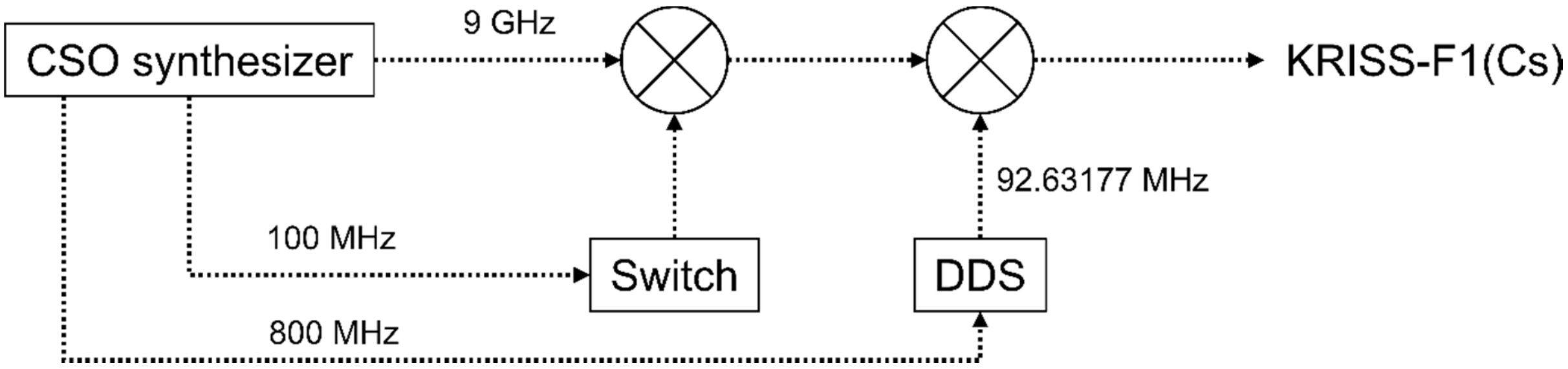

Fig. 4. Block diagram for the local oscillator for KRISS-F1(Cs) from the 9 GHz synthesizer output. Its pulsed operation is carried out by switching 100 MHz before the first mixer. Mixers are single-sideband mixers to suppress the unwanted carrier signals.

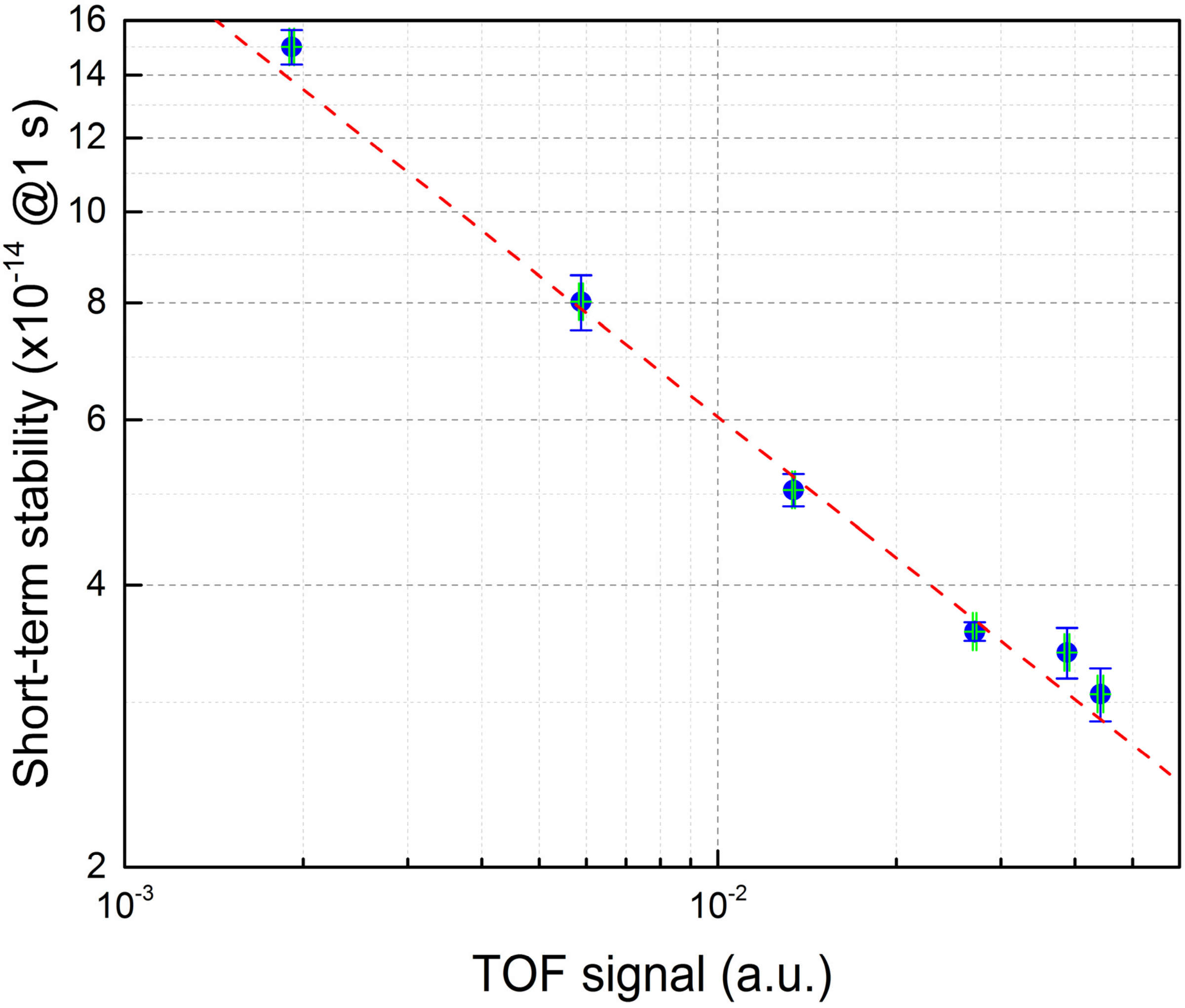

Fig. 5. Short-term stability measured by the KRISS-F1(Cs) with varying trapped atom number [11]. Dashed line is the fit according to (1). The horizontal axis represents the magnitude of the time-of-flight (TOF) signal which is proportional to the number of atoms.

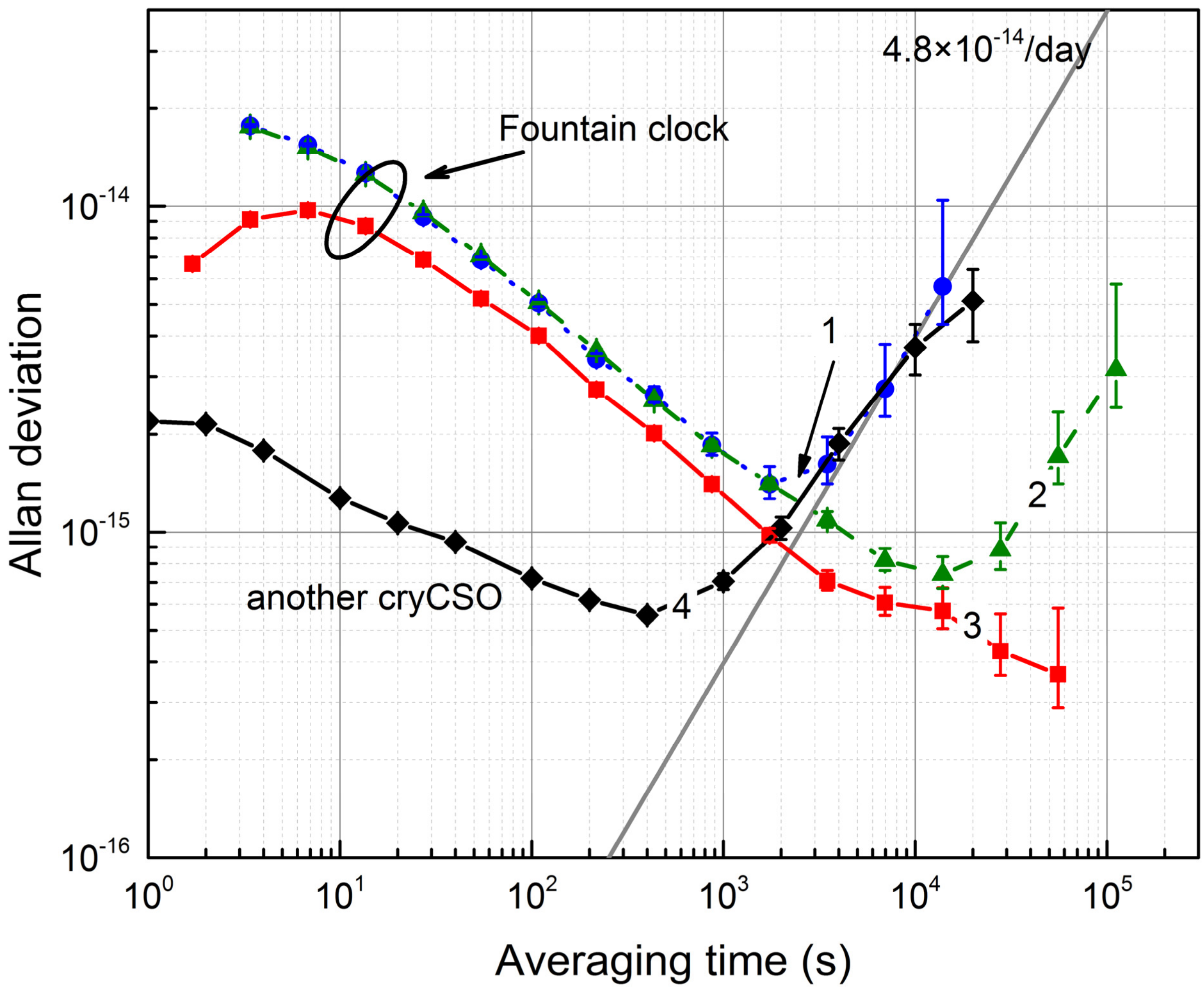

Fig. 6. The long-term stability of 9 GHz (curves 1, 2 and 3 with no drift compensation, the 1st compensation and the 2nd compensation, respectively) measured with the KRISS-F1(Cs) and the cryoCSO itself (curve 4) measured with another cryoCSO by Hartnett's group before the shipment to our lab. The grey line corresponds to the linear frequency drift of 4.8×10-14/day. Overall, the curve 3 shows better stability than the curves 1 and 2 because the curve 3 was measured with higher number of atoms.